\documentclass[prb,preprint,amsmath,amssymb,superscriptaddress,floatfix]{revtex4}
\usepackage{epsfig}
\usepackage{graphicx}
\usepackage{amsmath,amssymb}
\usepackage{bm}

\def\Rea{\mathrm{Re}\,}
\def\Ima{\mathrm{Im}\,}

\begin{document}

\title{Observation of a Berry phase anti-damping spin-orbit torque}
\author{H.~Kurebayashi}
\affiliation{Microelectronics Group, Cavendish Laboratory, University of Cambridge, CB3 0HE, UK}
\affiliation{PRESTO, Japan Science and Technology Agency, Kawaguchi 332-0012, Japan}
\author{Jairo Sinova}
\affiliation{Department of Physics, Texas A\&M
University, College Station, Texas 77843-4242, USA}
\affiliation{Institute of Physics ASCR, v.v.i., Cukrovarnick\'a 10, 162 53 Praha 6, Czech Republic}
\author{D.~Fang}
\author{A.~C.~Irvine}
\affiliation{Microelectronics Group, Cavendish Laboratory, University of Cambridge, CB3 0HE, UK}
\author{J.~Wunderlich}
\affiliation{Institute of Physics ASCR, v.v.i., Cukrovarnick\'a 10, 162 53 Praha 6, Czech Republic}
\affiliation{Hitachi Cambridge Laboratory, Cambridge CB3 0HE, UK}
\author{V.~Nov\'ak}
\affiliation{Institute of Physics ASCR, v.v.i., Cukrovarnick\'a 10, 162 53 Praha 6, Czech Republic}
\author{R.~P.~Campion}
\affiliation{School of Physics and
Astronomy, University of Nottingham, Nottingham NG7 2RD, UK}
\author{B.~L.~Gallagher}
\affiliation{School of Physics and
Astronomy, University of Nottingham, Nottingham NG7 2RD, UK}
\author{E.~K.~Vehstedt}
\affiliation{Department of Physics, Texas A\&M
University, College Station, Texas 77843-4242, USA}
\affiliation{Institute of Physics ASCR, v.v.i., Cukrovarnick\'a 10, 162 53 Praha 6, Czech Republic}
\author{L.~P.~Z\^ arbo}
\affiliation{Institute of Physics ASCR, v.v.i., Cukrovarnick\'a 10,
162 53 Praha 6, Czech Republic}
\author{K.~V\'yborn\'y}
\affiliation{Institute of Physics ASCR, v.v.i., Cukrovarnick\'a 10, 162 53 Praha 6, Czech Republic}
\author{A.~J.~Ferguson}
\affiliation{Microelectronics Group, Cavendish Laboratory, University of Cambridge, CB3 0HE, UK}
\author{T.~Jungwirth}
\affiliation{Institute of Physics ASCR, v.v.i., Cukrovarnick\'a 10, 162 53 Praha 6, Czech Republic}
\affiliation{School of Physics and Astronomy, University of Nottingham, Nottingham NG7 2RD, UK}
\date{\today}

\maketitle
\textbf{
Recent observations of  current-induced magnetization switching at ferromagnet/normal-conductor interfaces\cite{Miron:2011_b,Liu:2012_a}
have important consequences for future magnetic memory technology. In one interpretation, the switching originates from carriers with spin-dependent scattering giving rise to a relativistic anti-damping spin-orbit torque (SOT)\cite{Kim:2012_a,Pesin:2012_a,Wang:2012_b,Manchon:2012_a,Bijl:2012_a} in structures with broken space-inversion symmetry.\cite{Edelstein:1990_a,Manchon:2008_a,Manchon:2009_a,Chernyshov:2009_a,Garate:2009_a,Endo:2010_a,Fang:2010_a,Miron:2010_a,Pi:2010_a,Miron:2011_a,Miron:2011_b,Suzuki:2011_a,Kim:2012_a,Pesin:2012_a,Wang:2012_b,Manchon:2012_a,Bijl:2012_a,Kim:2012_b,Garello:2013_a} The alternative interpretation  \cite{Miron:2011_b,Liu:2011_b,Liu:2012_a,Pesin:2012_a,Wang:2012_b,Manchon:2012_a,Kim:2012_b,Garello:2013_a} combines the relativistic spin Hall effect (SHE),\cite{Dyakonov:1971_b,Hirsch:1999_a,Murakami:2003_a,Sinova:2004_a,Kato:2004_d,Wunderlich:2004_a,Jungwirth:2012_a} making the normal-conductor  an injector of a spin-current, with the non-relativistic spin-transfer torque (STT)\cite{Slonczewski:1996_a,Berger:1996_a,Ralph:2008_a,Brataas:2012_a} in the ferromagnet. Remarkably, the SHE in these experiments originates from the Berry phase effect in the band structure of a clean crystal\cite{Murakami:2003_a,Sinova:2004_a,Wunderlich:2004_a,Tanaka:2007_b,Liu:2012_a} and the anti-damping STT is also based on a disorder-independent transfer of spin  from carriers to magnetization. Here we report the observation of an anti-damping SOT stemming from an analogous Berry phase effect to the SHE. The SOT alone can therefore induce magnetization dynamics based on a scattering-independent principle. The ferromagnetic semiconductor (Ga,Mn)As we use has a broken space-inversion symmetry in the crystal.\cite{Chernyshov:2009_a,Garate:2009_a,Endo:2010_a,Fang:2010_a} This allows us to consider a bare ferromagnetic film which eliminates by design any SHE related contribution to the spin torque. We provide an intuitive picture of the Berry phase origin of the anti-damping SOT and a microscopic modeling of measured data.}

In the quasiclassical transport theory, the linear response of the carrier system to the applied electric field is described by the non-equilibrium  distribution function of carrier eigenstates which are considered to be unperturbed by the electric field. The form of the non-equilibrium distribution function is obtained by accounting for the combined effects of the carrier acceleration in the field and scattering. On the other hand, in the time-dependent quantum-mechanical perturbation theory the linear response is described by  the equilibrium distribution function and  by the perturbation of carrier wavefunctions in the applied electric field. The latter framework was the basis of the intrinsic Berry phase mechanism introduced to explain the anomalous Hall effect in (Ga,Mn)As\cite{Jungwirth:2002_a} and, subsequently, in a number of other ferromagnets.\cite{Nagaosa:2010_a} Via the anomalous Hall effect, the Berry phase physics entered the field of the SHE in spin-orbit coupled paramagnets. Here the concept of a scattering-independent origin brought the attention of a wide physical community to this relativistic phenomenon, eventually turning the SHE into an important field of condensed matter physics and spintronics.\cite{Jungwirth:2012_a}  In our work we demonstrate that the SOT can also have the relativistic quantum-mechanical Berry phase origin.

We start by deriving the intuitive picture of our Berry phase anti-damping SOT based on the Bloch equation description of the carrier spin dynamics. In (Ga,Mn)As,  the combination of the broken inversion symmetry of the zinc-blende lattice and strain can produce spin-orbit coupling terms in the Hamiltonian which are linear in momentum and have the Rashba symmetry, $H_{R}=\frac{\alpha}{\hbar}(\sigma_xp_y-\sigma_yp_x)$, or the Dresselhaus symmetry, $H_{D}=\frac{\beta}{\hbar}(\sigma_xp_x-\sigma_yp_y)$ (see Fig.~1a).\cite{Chernyshov:2009_a,Garate:2009_a,Endo:2010_a,Fang:2010_a} Here $\bm{\sigma}$ are the Pauli spin matrices and $\alpha$ and $\beta$ represent the strength of the Rashba and Dresselhaus spin-orbit coupling, respectively. The interaction between carrier spins and magnetization is described by the exchange Hamiltonian term, $H_{ex}=J\bm{\sigma}\cdot\mathbf{M}$. In (Ga,Mn)As,  $\mathbf{M}$ corresponds to the ferromagnetically ordered local moments on the Mn $d$-orbitals and $J$ is the antiferromagnetic carrier--local moment kinetic-exchange constant.\cite{Jungwirth:2006_a}  The physical origin of our anti-damping SOT is best illustrated assuming for simplicity a 2D parabolic form of the spin-independent part of the total Hamiltonian, $H=\frac{p^2}{2m}+H_{R(D)}+H_{ex}$, and the limit of $H_{ex}\gg H_{R(D)}$. In equilibrium, the carrier spins are then approximately aligned with the exchange field, independent of their momentum. The origin of the SOT can be understood from solving the Bloch equations for carrier spins during the acceleration of the carriers in the applied electric field, i.e., between the scattering events. Let's define $x$-direction as the direction of the applied electric field $\mathbf{E}$. For $-\mathbf{M}\parallel\mathbf{E}$, the equilibrium effective magnetic field acting on the carrier spins, $\mathbf{s}=\frac{\bm{\sigma}}{2}$,  due to the exchange term is, $\mathbf{B}^{eq}_{eff}\approx(2JM,0,0)$, in units of energy. During the acceleration in the applied electric field, $\frac{dp_x}{dt}=eE_x$, and the effective magnetic field acquires a time-dependent $y$-component due to $H_R$ for which $\frac{d B_{eff,y}}{dt}=\frac{2\alpha}{\hbar}\frac{dp_x}{dt}$, as illustrated in Fig.~1b. For small tilts of the spins from equilibrium,  the Bloch equations $\frac{d\mathbf{s}}{dt}=\frac{1}{\hbar}(\mathbf{s}\times \mathbf{B}_{eff})$ yield, $s_x\approx s$, $s_y\approx  s\frac{B_{eff,y}}{B^{eq}_{eff}}$, and
\begin{equation}
s_z\approx -\frac{\hbar s}{(B^{eq}_{eff})^2}\frac{dB_{eff,y}}{dt}=-\frac{s}{2J^2 M^2}\alpha eE_x\;.
\label{s_zM_x}
\end{equation}
The non-equilibrium spin orientation of the carries acquires a time and momentum independent $s_z$ component.

As illustrated in Figs.~1b,c, $s_z$ depends on the direction of the magnetization $\mathbf{M}$ with respect to the applied electric field. It has a maximum for $\mathbf{M}$ (anti)parallel to $\mathbf{E}$ and vanishes for $\mathbf{M}$ perpendicular to $\mathbf{E}$. For a general direction of $\mathbf{M}$ we obtain,
\begin{equation}
s_{z,\mathbf{M}}\approx \frac{s}{2J^2 M^2}\alpha eE_x\cos\theta_{\mathbf{M-E}}\;.
\label{s_zM}
\end{equation}
The total non-equilibrium spin polarization $S_z$ is obtained by integrating $s_{z,\mathbf{M}}$ over all occupied states. The non-equilibrium spin polarization produces an out-of-plane field which exerts a torque on the in-plane magnetization. From Eq.~(\ref{s_zM}) we obtain that this intrinsic SOT is anti-damping-like,
\begin{equation}
\frac{d\mathbf{M}}{dt}=\frac{J}{\hbar}(\mathbf{M}\times S_z\hat{z})\sim \mathbf{M}\times([\mathbf{E}\times\hat{z}]\times\mathbf{M})\,.
\label{SOT_R}
\end{equation}
For the Rashba spin-orbit coupling, Eq.~(\ref{SOT_R}) applies to all directions of the applied electric field with respect to crystal axes.  By replacing $H_R$ with $H_D$ we can follow the same arguments and arrive at the corresponding expressions for the anti-damping  SOT. In the case of the Dresselhaus spin-orbit coupling, the symmetry of the anti-damping SOT depends on the direction of $\mathbf{E}$ with respect to crystal axes, as seen from Fig.~1a. For a particular electric field direction one can interpolate the angle, $\theta_{\mathbf{M-E}}$, dependence of
the Dresselhaus  out of plane field by the relative phase of the Rashba and Dresselhaus polarization along $\mathbf{E}$.
In the following table we summarise the angle dependence of the Rashba and Dresselhaus contributions to $s_{z,\mathbf{M}}$ for electric fields along different crystal directions.

\begin{center}
\begin{tabular}{|c|c|c|}
  \hline
   & Rashba: $s_{z,\mathbf{M}}\sim$ & Dresselhaus: $s_{z,\mathbf{M}}\sim$ \\ \hline
  $\mathbf{E}\parallel[100]$ & $\cos{\theta_{\mathbf{M-E}}}$ & $\sin{\theta_{\mathbf{M-E}}}$ \\ \hline
  $\mathbf{E}\parallel[010]$ & $\cos{\theta_{\mathbf{M-E}}}$ & $-\sin{\theta_{\mathbf{M-E}}}$ \\ \hline
  $\mathbf{E}\parallel[110]$ & $\cos{\theta_{\mathbf{M-E}}}$ & $\cos{\theta_{\mathbf{M-E}}}$ \\ \hline
  $\mathbf{E}\parallel[1-10]$ & $\cos{\theta_{\mathbf{M-E}}}$ & $-\cos{\theta_{\mathbf{M-E}}}$ \\
  \hline
\end{tabular}
\end{center}

To highlight the analogy between our anti-damping SOT and the Berry phase origin of the SHE\cite{Murakami:2003_a,Sinova:2004_a} we illustrate in Fig.~1d the solution of the Bloch equations in the absence of the exchange Hamiltonian term.\cite{Sinova:2004_a} In this case  $\mathbf{B}^{eq}_{eff}$ depends on the carrier momentum which implies a momentum-dependent $z$-component of the non-equilibrium spin,
\begin{equation}
s_{z,\mathbf{p}}\approx \frac{s\hbar^2}{2\alpha p^2}\alpha eE_x\sin\theta_{\mathbf{p}}\;.
\label{s_zp}
\end{equation}
Clearly the same spin rotation mechanism which generates the spin accumulation in the case of our anti-damping SOT (Fig.~1b) is responsible for the scattering-independent spin-current in the SHE (Fig.~1d).

To complete the picture of the common origin between the microscopic physics of the Berry phase SHE and our anti-damping SOT
we point out that equivalent expressions for the SHE  spin current 
and the SOT spin polarization
can be obtained from the quantum-transport Kubo formula.
The expression for the out-of-plane non-equilibrium spin polarization that generates our
anti-damping SOT is given by
\begin{equation}
S_z=\frac{\hbar}{V}\sum_{\mathbf{k},a\neq b}(f_{\mathbf{k},a}-f_{\mathbf{k},b})\frac{{\rm Im}[\langle\mathbf{k},a|s_z|\mathbf{k},b\rangle\langle\mathbf{k},b|e\bm E\cdot\mathbf{v}|\mathbf{k},a\rangle]}{(E_{\mathbf{k},a}-E_{\mathbf{k},b})^2}\,,
\label{Kubo:clean}
\end{equation}
where $a,b$ indicate the band indices. Here $f_{\mathbf{k}a}$ are the Fermi-Dirac distribution functions corresponding to band
energies $E_{\mathbf{k}a}$.
This expression is analogous to Eq. (9) in Ref. \onlinecite{Sinova:2004_a} for the Berry phase intrinsic SHE.

We now discuss our low-temperature (6~K) experiments in which we identify the presence of the anti-damping SOT in our in-plane magnetized (Ga,Mn)As samples. We follow the methodology of several previous experiments\cite{Liu:2012_a,Fang:2010_a} and use current induced ferromagnetic resonance to investigate the magnitude and symmetries of the alternating fields responsible for resonantly driving the magnetisation. In our experiment, illustrated schematically in Fig.~\ref{fig2}a, a signal generator drives a microwave frequency current through a $4~\mu m\times40~\mu m$ micro-bar patterned from a 18 nm thick (Ga,Mn)As epilayer with nominal 5\% Mn-doping. A bias tee is used to measure the dc voltage across the sample, which is generated according to Ohm's law due to the product of the oscillating magneto-resistance (during magnetisation precession) and the microwave current.\cite{Tulapurkar:2005_a} Solving the equation of motion for the magnetisation (the LLG equation) for a small excitation field vector $(h_x,h_y,h_z)\exp{[i\omega t]}$ we find dc voltages containing symmetric ($V_S$) and anti-symmetric ($V_A$) Lorentzian functions, shown in Fig.~\ref{fig2}b. As the saturated magnetization of the sample is rotated, using $\theta_{\mathbf{M-E}}$ to indicate the angle from the current/bar direction, the in-plane and out-of plane components
of the excitation field are associated with $V_S$ and $V_A$ via:
\begin{equation}
  V_S\propto h_z \sin{2\theta_{\mathbf{M-E}}}\,\,,
\end{equation}
\begin{equation}
  V_A\propto -h_x\sin{\theta_{\mathbf{M-E}}}\sin{2\theta_{\mathbf{M-E}}}+h_y \cos{\theta_{\mathbf{M-E}}}\sin{2\theta_{\mathbf{M-E}}}\,\,.
\end{equation}

In this way we are able to determine, at a given magnetization orientation, the current induced field vector. In Fig.~\ref{fig2}c we show the angle dependence of $V_S$ and $V_A$ for an in-plane rotation of the magnetization for a micro-bar patterned in the [100] crystal direction. As described in the Supplementary information, the voltages $V_S$ and $V_A$ are related to the alternating excitation field, using the micro-magnetic parameters and AMR of the sample. The in-plane field components, determined from $V_A$, are well fitted by a M-independent current induced field vector ($\mu_0h_x,\mu_0h_y)$=(-91,-15)$~\mu$T referenced to a current density of $10^5~$Acm$^{-2}$. Since $V_S$ is non-zero, it is seen that there is also a significant $h_z$ component of the current induced field. Furthermore, since $V_S$ is not simply described by $\sin{2\theta_{\mathbf{M-E}}}$, $h_z$ depends on the in-plane orientation of the magnetization. To analyse the symmetry of the out-of-plane field we fit the angle dependence of $V_S$, finding for the [100] bar shown in (Fig.~\ref{fig2}c) that $\mu_0h_z=(13 + 95\sin{\theta_{\mathbf{M-E}}}+41\cos{\theta_{\mathbf{M-E}}})~\mu$T.

We measure 8 samples, 2 patterned in each crystal direction and plot in Fig.~\ref{fig3} the resulting $\sin{\theta_{\mathbf{M-E}}}$ and $\cos{\theta_{\mathbf{M-E}}}$ coefficients  of $h_z$. The corresponding in-plane fields are also shown: since these are approximately magnetisation-independent they can be represented in Fig.~\ref{fig3} by a single vector. In the [100] bar we found that the $\sin{\theta_{\mathbf{M-E}}}$ coefficient of $h_z$, which according to the theoretical model originates in the Dresselhaus spin-orbit term, is greater than the $\cos{\theta_{\mathbf{M-E}}}$ coefficient related to the Rashba spin-orbit term (see Table 1). If we examine the symmetries of $h_z$ in our sample set, we find that they change in the manner expected for samples with dominant Dresselhaus term; a trend that is in agreement with the in-plane fields. The angle-dependence of $h_z$ measured throughout our samples indicates an anti-damping like SOT with the theoretically predicted symmetries. Since the magnitude of the measured $h_z$ is comparable to the in-plane fields (see Supplementary information for a detailed comparison), the anti-damping and field-like SOTs are equally important for driving the magnetisation dynamics in our experiment.

To model the measured anti-damping SOT, assuming  its Berry phase intrinsic origin, we start from the effective kinetic-exchange Hamiltonian describing (Ga,Mn)As:\cite{Jungwirth:2006_a}
$
  H=H_{\rm KL}+H_{\rm strain}+H_{\rm ex}$. Here $H_{\rm ex}=J_{\rm pd}c_{\rm Mn} S_{\rm Mn} \hat{\mathbf{M}}\cdot{\mathbf s}$,
$H_{\rm KL}$ and $H_{\rm strain}$ refer to the strained Kohn-Luttinger Hamiltonian for the hole systems of GaAs (see Supplementary information),
$\mathbf s$ is the hole spin operator, $S_{\rm Mn}=5/2$, $c_{\rm Mn}$ is the
Mn density, and $J_{\rm pd}=55$ meV nm$^{3}$ is the kinetic-exchange coupling between the localized $d$-electrons and the
valence band holes.
The Dresselhaus and Rashba symmetry parts of the strain Hamiltonian in the hole-picture are given by
\begin{eqnarray}
  \label{eq:c4strain_main}
  H_{\rm strain}
&=& -    3C_4\left[
    s_x\left(\epsilon_{yy}-\epsilon_{zz}\right)k_x+{\rm c.p.}
  \right] \\
&& -  3C_5
  \left[
    \epsilon_{xy}(k_ys_x-k_xs_y)+{\rm c.p.}
  \right] ,\nonumber
\end{eqnarray}
where $C_4 = 10$ eV${\rm \AA}$ and we take $C_5=C_4$.\cite{Silver:1992_a,Stefanowicz:2010_a}
These momentum-dependent $H_{\rm strain}$ terms are
 essential for the generation of  SOT because they break the space-inversion symmetry.
The momentum-dependent spin-orbit contribution to $H_{\rm KL}$ does not produce directly a SOT but
it does interfere with the linear in-plane momentum terms in $H_{\rm strain}$ to reduce the magnitude of the SOT and introduce higher harmonics in the $\theta_{\mathbf{M-E}}$ dependence of $\mu_0h_z$. We have also replaced $H_{\rm KL}$ with a parabolic model with effective mass $m^{\ast}=0.5 m_e$
and included the spin-orbit coupling only through the Rashba and Dresselhaus-symmetry  strain terms given by Eq.~(\ref{eq:c4strain_main}).
The expected $\cos{\theta_{\mathbf{M-E}}}$ or $\sin{\theta_{\mathbf{M-E}}}$  symmetry without higher harmonics follows. In addition, a large increase of the amplitude of the effect is observed since the broken inversion symmetry spin-texture does not compete with the centro-symmetric one induced by the large spin-orbit coupled $H_{\rm KL}$ term. This indicates that for a system in which the dominant spin-orbit coupling  is linear in momentum our Berry phase anti-damping SOT will be largest.

In  Fig.~\ref{fig4} we show calculations for our (Ga,Mn)As samples including the spin-orbit coupled $H_{\rm KL}$ term (full lines) term or replacing it with the parabolic model (dashed lines).
The non-equilibrium spin density induced by the Berry phase effect is obtained from
the Kubo formula:\cite{Garate:2009_a}
\begin{equation}\label{eq:kubo_cif_main}
   S_z = \frac{\hbar }{2\pi V} \Rea \sum_{\mathbf{k},a\ne b}
   \langle\mathbf{k},a|{s_z}|\mathbf{k},b\rangle\langle\mathbf{k},b|e\bm E\cdot\mathbf{v}|\mathbf{k},a\rangle
   [G^A_{\mathbf{k}a}G^R_{\mathbf{k}b}-G^R_{\mathbf{k}a}G^R_{\mathbf{k}b}],
\end{equation}
where the Green's functions
$G^R_{\mathbf{k}a}(E)|_{E=E_F}\equiv G^R_{\mathbf{k}a}=1/(E_F-E_{\mathbf{k}a}+i\Gamma)$,
with the property $G^A=(G^R)^*$. $E_F$ is the Fermi energy and $\Gamma$ is the disorder induced spectral broadening, taken in the simulations
to be 25 meV. Note that in the disorder-free limit, Eq.~(\ref{eq:kubo_cif_main}) turns into Eq.~(\ref{Kubo:clean}) introduced above.
The relation between $S_z$ and the  effective magnetic field generating the Berry phase SOT is given by $\mu_0 h_z=-({J_{\rm pd}}/{g\mu_{\rm B}})S_z$,
where $\mu_{\rm B}$ is the Bohr magneton, and $g=2$
corresponds to the localized  $d$-electrons in (Ga,Mn)As (for more details on the modeling see Supplementary Information).

Results of our calculations are compared in Fig.~\ref{fig4} with experimental dependencies of $h_z$ on $\theta_{\mathbf{M-E}}$ measured in the 4 micro-bar directions. As expected, the parabolic model calculations strongly overestimate the SOT field $h_z$. On the other hand, including the competing centro-symmetric $H_{\rm KL}$ term, which is present in the (Ga,Mn)As valence band, gives the correct order of magnitude of  $h_z$ as compared to experiment. Moreover, by including the $H_{\rm KL}$ term we can also explain the presence of higher harmonics in the $\theta_{\mathbf{M-E}}$ dependencies seen in experiment. This confirms that the experimentally observed anti-damping SOT is of the Berry phase origin.

To conclude, we have predicted a Berry phase SOT phenomenon and experimentally identified the effect in (Ga,Mn)As which is a model ferromagnetic system with broken space inversion symmetry in the bulk crystal.  Learning from the analogy with the intrinsic Berry phase anomalous Hall effect, first identified in (Ga,Mn)As and subsequently observed in a number of ferromagnets, we infer that our Berry phase SOT is a generic phenomenon in spin-orbit coupled magnetic systems with broken space-inversion symmetry. In particular, the Berry phase SOT might be present in ferromagnet/paramagnet bilayers with the  broken structural inversion symmetry. The resulting Rashba-like anti-damping SOT has the same basic symmetry of its magnetization dependence as the earlier reported SHE-STT mechanism. Therefore, two strong relativistic mechanisms of scattering-independent origin can contribute in the current-induced magnetization switching in these technologically important magnetic structures.

\section*{Methods and Materials}
\textbf{Materials:} The 18 nm thick (Ga$_{0.95}$,Mn$_{0.05}$)As epilayer was grown on a GaAs [001] substrate by molecular beam epitaxy, performed at a substrate temperature of 230 C. It was subsequently annealed for 8 hours at 200 C. It has a Curie temperature of 132 K; a room temperature conductivity of 387 $\Omega^{-1}{\rm cm}^{-1}$ which increases to 549 $\Omega^{-1}{\rm cm}^{-1}$ at 5 K; and has a carrier concentration at 5 K determined by high magnetic field Hall measurement of $1.1 \times 10^{21}$~${\rm cm}^{-3}$.

\textbf{Devices:} Two terminal microbars are patterned in different crystal directions by electron beam lithography to have dimensions of $4 \times 40$~$\mu m$. These bars have a typical low temperature resistance of 10 $k\Omega$ (data-table in supplementary information).

\textbf{Experimental procedure:} A pulse modulated (at 789 Hz) microwave signal (at 11 GHz) with a source power of (20 dBm) is transmitted down to cryogenic temperatures using low-loss, low semi-rigid cables. The microwave signal is launched onto a printed circuit board patterned with a coplanar waveguide, and then injected into the sample via a bond-wire. The rectification voltage, generated during microwave precession, is separated from the microwave circuit using a bias tee, amplified with a voltage amplifier and then detected with lock-in amplifier. All measurements are performed with the samples at 6 K.

\textbf{Calibration of microwave current:} The resistance of a (Ga,Mn)As micro-bar depends on temperature, and therefore on the Joule heating by an electrical current. First, the resistance change of the micro-bar due to Joule heating of a direct current is measured. Then, the resistance change is measured as a function of applied microwave power. We assume the same Joule heating (and therefore resistance change of the micro-bar) for the same direct and rms microwave currents, enabling us to calibrate the unknown microwave current against the known direct current.

\section*{Corresponding author}
Correspondence and requests for materials should be addressed to AJF. (ajf1006@cam.ac.uk).
\section*{Acknowledgment}
We acknowledge fruitful discussions with, and support from EU ERC Advanced Grant No. 268066, EU grant FP7 215368 SemiSpinNet, from the Ministry of Education of the Czech Republic Grant No. LM2011026, from the Academy of Sciences of the Czech Republic No. AV0Z10100521, Praemium Academiae,
and from U.S. grants ONR-N000141110780, NSF-DMR-1105512 and NSF TAMUS LSAMP BTD Award 1026774. AJF acknowledges support from a Hitachi research fellowship. HK acknowledges financial support from the JST.
\section*{Author contributions}
Theory and data modeling: TJ, EKV, LPZ, KV, JS; Materials: VN, RPC, BLG; Sample preparation: ACI; experiments and data analysis: HK, DF, JW, AJF; writing: TJ, AJF, HK, JS;  project planning: TJ, AJF, JS.

\begin{figure}[h!]
\vspace*{0cm}
\includegraphics[width=.9\columnwidth,angle=0]{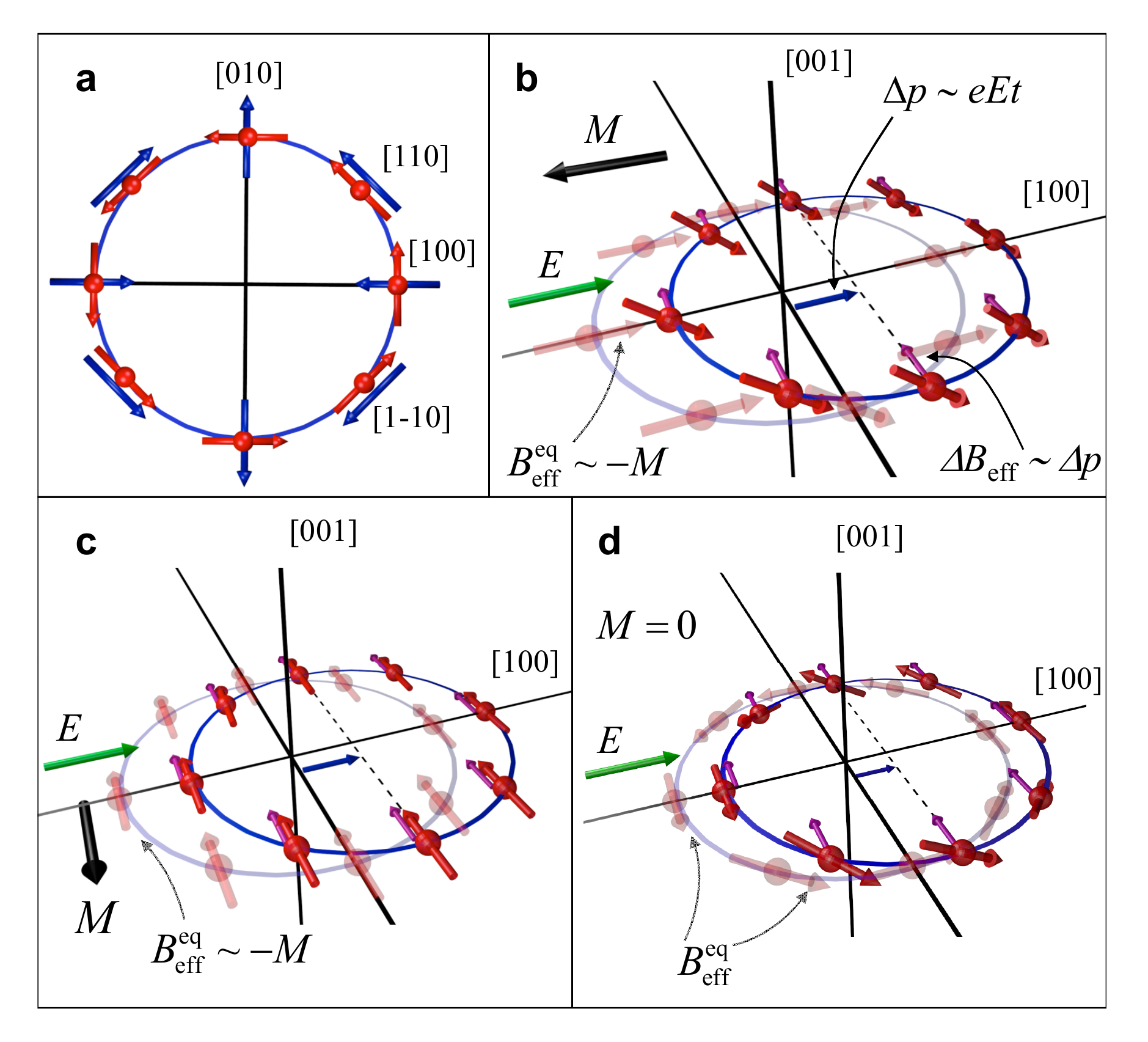}
\vspace*{-0cm}
\caption{\textbf{Spin-orbit coupling and anti-damping SOT.} {\bf a,} Rashba (red) and Dresselhaus (blue) spin textures.
{\bf b,} For the case of a Rashba-like symmetry, the out-of-plane non-equilibrium carrier spin-density that generates the Berry phase anti-damping SOT has a maximum for $\mathbf{E}$ (anti)parallel to $\mathbf{M}$. In this configuration the equilibrium effective field $B^{eq}_{eff}$ and the additional field $\Delta B_{eff}\perp \mathbf{M}$ due to the acceleration are perpendicular to each other causing all spins to tilt in the same out-of-plane direction.
{\bf c,} For the case of a Rashba-like symmetry, the out-of-plane non-equilibrium carrier spin-density is zero for $\mathbf{E}\perp \mathbf{M}$ since $B^{eq}_{eff}$ and $\Delta B_{eff}$ are parallel to each other.
{\bf d,}  The analogous physical phenomena for zero magnetization induces a tilt of the spin out of the plane that has opposite sign for momenta pointing to the
left or the right of the electric field, inducing in this way the intrinsic Berry phase SHE.\cite{Sinova:2004_a}}
\label{fig1}
\end{figure}

\begin{figure}[h!]
\vspace*{0cm}
\includegraphics[width=.5\columnwidth,angle=0]{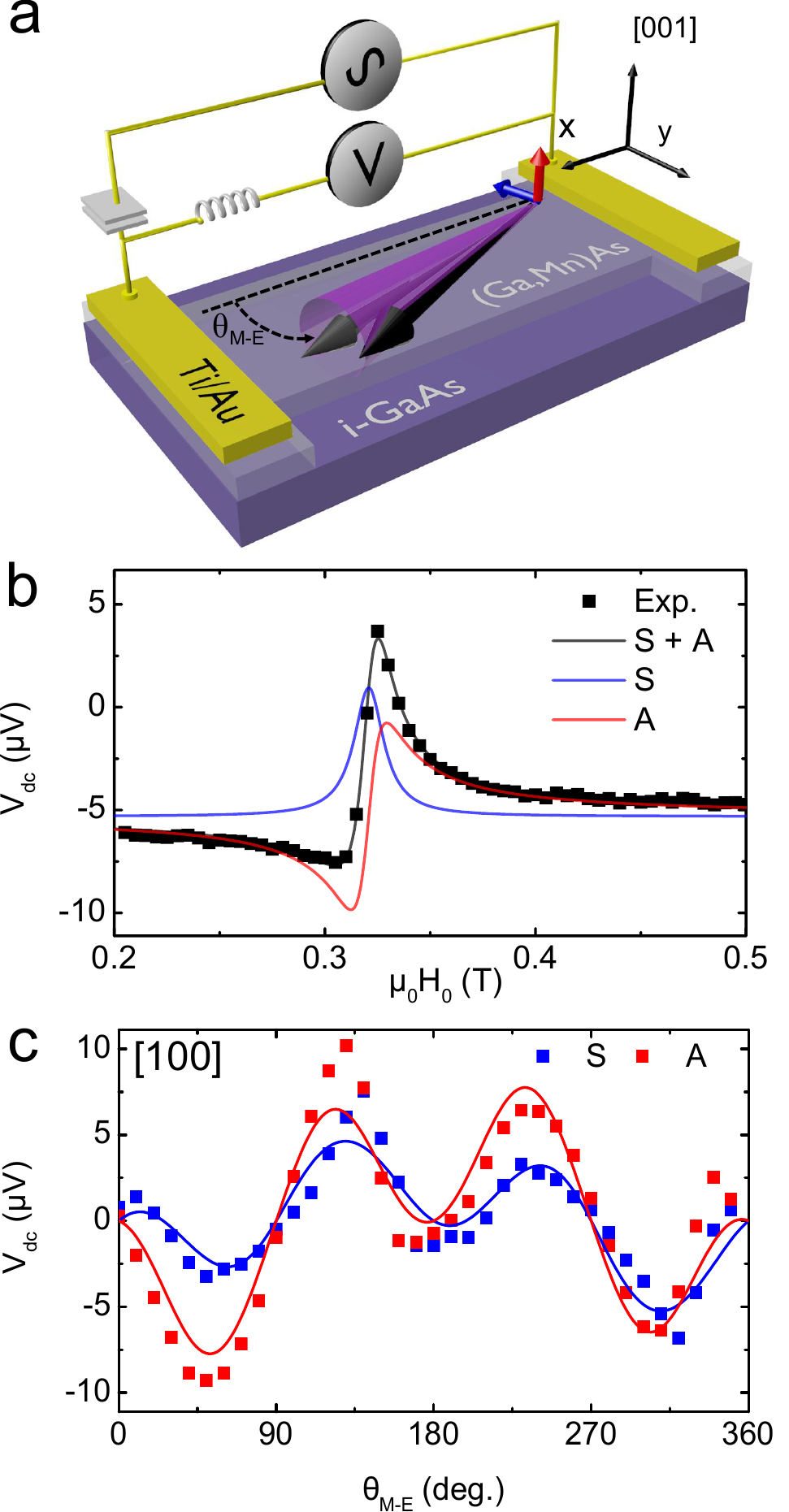}
\vspace*{-0cm}
\caption{\textbf{Spin-orbit FMR experiment.} {\bf a,} Schematic of the sample, measurement setup and magnetisation precession. Microwave power goes through a bias-tee and into the (Ga,Mn)As micro-bar which is placed inside a cryostat. The injected microwave current drives FMR that is detected via a dc voltage $V_{\rm dc}$ across the micro-bar. We define $\theta_{\mathbf{M}-\mathbf{E}}$ as an angle of the static magnetisation direction determined by the external magnetic field, measured from the current flow direction.
{\bf b,} Typical spin-orbit FMR signal driven by an alternating current at 11~GHz and measured by $V_{\rm dc}$ as a function of external magnetic field. The data were fitted by a combination of symmetric and anti-symmetric Lorentzian functions.
{\bf c,} Symmetric and antisymmetric component of $V_{\rm dc}$
as a function of $\theta_{\mathbf{M}-\mathbf{E}}$ for current along the $[100]$ direction.
}
\label{fig2}
\end{figure}

\begin{figure}[h!]
\vspace*{0cm}
\includegraphics[width=.9\columnwidth,angle=0]{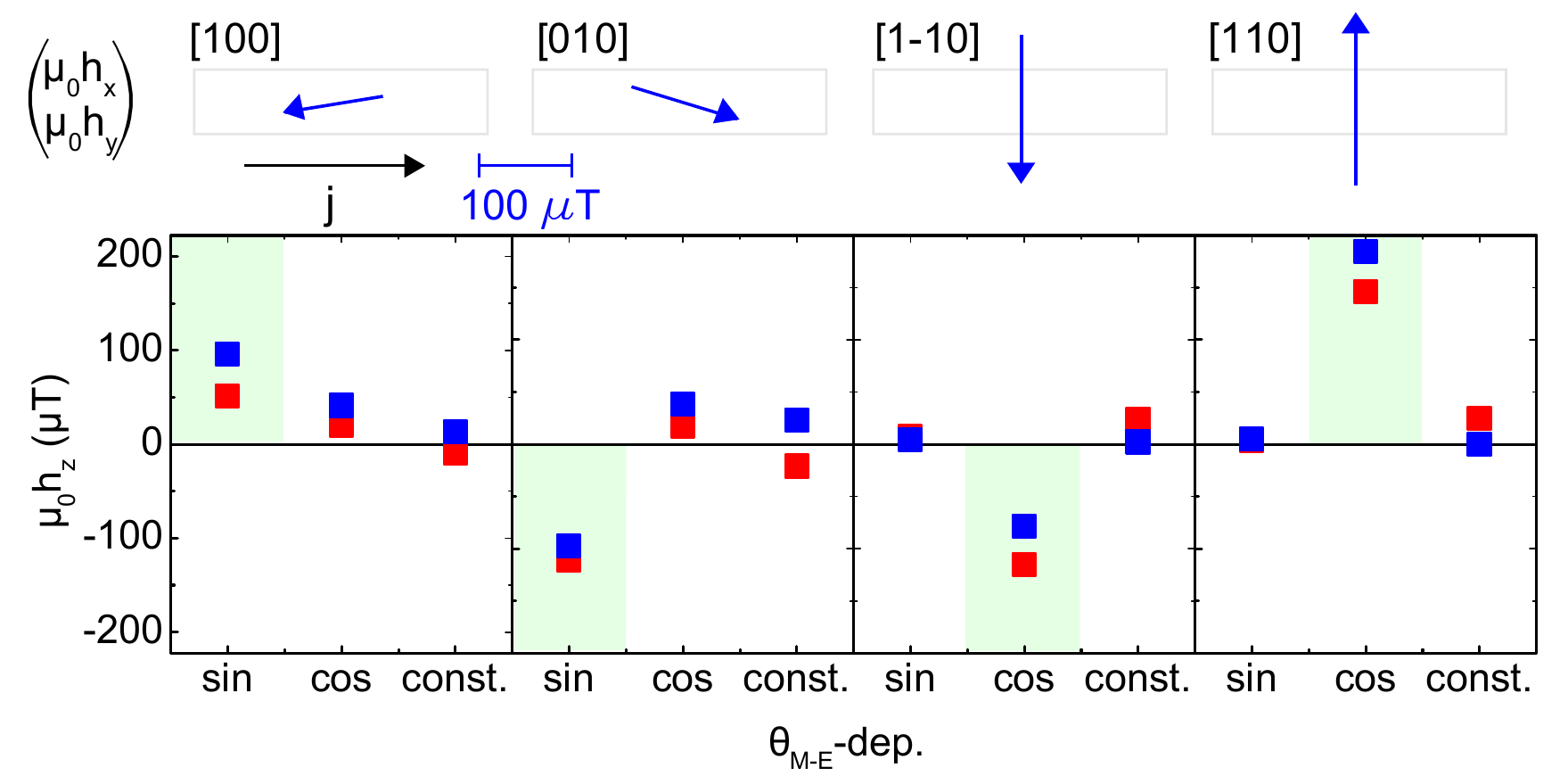}
\vspace*{-0cm}
\caption{\textbf{In-plane and out-of-plane SOT fields.} In-plane spin-orbit field and coefficients of the $\cos{\theta_{\mathbf{M-E}}}$ and $\sin{\theta_{\mathbf{M-E}}}$ fits to the angle-dependence of out-of-plane SOT field for our sample set. For the in-plane fields, a single sample in each micro-bar direction is shown (corresponding to the same samples that yield the blue out-of-plane data points). In the out-of-plane data, 2 samples are shown in each micro-bar direction. The symmetries expected for the anti-damping SOT, on the basis of the theoretical model for the Dresselhaus term in the spin-orbit interaction, are shown by light green shading. All data are normalised to a current density of $10^5$~Acm$^{-2}$.}
\label{fig3}
\end{figure}

\begin{figure}[h!]
\vspace*{0cm}
\includegraphics[width=.9\columnwidth,angle=0]{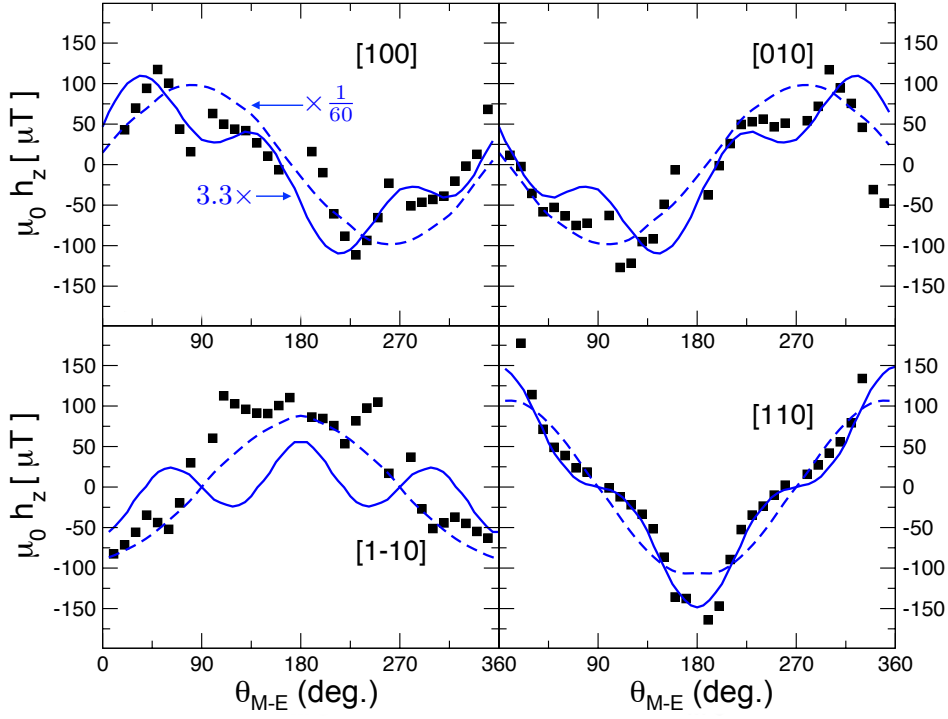}
\vspace*{-0cm}
\caption{\textbf{Theoretical modeling of the measured angular dependencies of the SOT fields.}
Microscopic model calculation for the measured (Ga,Mn)As samples assuming Rashba ($\epsilon_{xy}=-0.15\%$) and
Dresselhaus ($\epsilon_{xx}=-0.3\%$) strain. Solid blue lines correspond to the calculations with the centro-symmetric $H_{\rm KL}$ term included in the (Ga,Mn)As Hamiltonian. Dashed blue lines correspond to replacing $H_{\rm KL}$ with the parabolic model. Both calculations are done with a disorder broadening $\Gamma=25$ meV. Black points are experimental data whose fitting coefficients of the $\cos{\theta_{\mathbf{M-E}}}$ and $\sin{\theta_{\mathbf{M-E}}}$ first harmonics correspond to blue points in Fig.~3.}
\label{fig4}
\end{figure}

\clearpage

\begin{appendix}
\begin{center}
\section*{Supplementary Information}
\end{center}
\subsection{FMR linewidth analysis and sample parameters}
We use the phenomenological Landau-Lifshitz-Gilbert (LLG) equation to describe the spin-orbit-induced magnetisation dynamics in our (Ga,Mn)As micro-bars:
\begin{equation}
        \frac{\partial\mathbf{M}}{\partial t} = -\gamma \mathbf{M} \times \mathbf{H}_\text{tot} + \frac{\alpha}{M_s} \left( \mathbf{M} \times \frac{\partial\mathbf{M}}{\partial t}\right)-\gamma \mathbf{M} \times \mathbf{h}_\text{so}
            \label{eq:LLG}
    \end{equation}

\begin{figure}[b!]
\includegraphics[angle=0,width=0.50\textwidth]{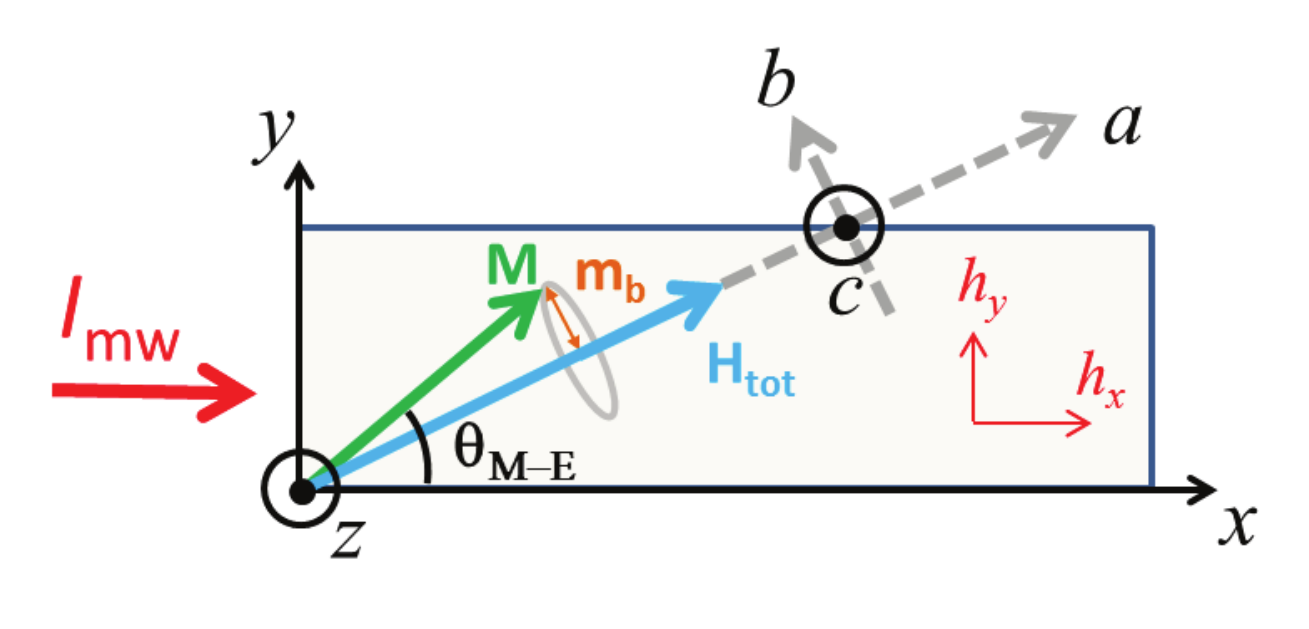}
\parbox{15cm}{\caption{The co-ordinate systems used. These are either defined with respect to the current direction (in the case of the spin orbit field) or with respect to the magnetisation (in the derivation of the rectification voltage).}\label{figS1}}
\end{figure}

Here, $\gamma$ and $\alpha$, M$_s$ are the gyromagnetic ratio, the dimensionless Gilbert damping constant and the saturation magnetisation respectively. The first term describes precession of the magnetisation $\bold{M}$ around the total static magnetic field $\bold{H_{\text{tot}}}$, which includes both magneto-crystalline anisotropy fields and the externally-applied field. Relaxation towards the equilibrium direction is expressed by the second term. When $\bold{M}$ is resonantly driven, in our case by the SO-torques included as the third term in the equation, it undergoes steady-state precession around the H$_{\text{tot}}$ direction. We assume a small precession angle, such that the magnetisation dynamics is within the linear excitation regime, hence we can write $\mathbf{M} = (M_s, m_{b} e^{i\omega t}, m_c e^{i\omega t})$ within the right-hand coordinate system defined by the equilibrium orientation of $\mathbf{M}$ (shown in Fig.~\ref{figS1}). In this coordinate system $\mathbf{h}_\text{so}$ can be given by the following, where $\theta_{\mathbf{M-E}}$ is the angle between $\mathbf{M}$ and the current direction.
 \begin{equation}
        \mathbf{h}_\text{so} = \left(
            {\begin{array} {c}
                h_x\cos\theta_{\mathbf{M-E}} + h_y\sin\theta_{\mathbf{M-E}}   \\
                -h_x\sin\theta_{\mathbf{M-E}} + h_y\cos\theta_{\mathbf{M-E}}   \\
                h_z
            \end{array}}    \right) e^{i\omega t},
            \label{hso}
    \end{equation}

Solving the LLG equation to first order, the expression for m$_b$ can be found as:
      \begin{equation}
        m_b = - \frac{[i(\omega / \gamma )h_z + (H_0 + H_1 + i\Delta H)( - h_x\sin\theta_{\mathbf{M-E}} + h_y\cos\theta_{\mathbf{M-E}})]M_{\text{s}}}{(\omega / \gamma)^2 - (H_0 + H_1 + i\Delta H)(H_0 + H_2 + i\Delta H) }
            \label{eq:mb}
    \end{equation}
    where $\Delta H = \alpha\omega/\gamma$ and $H_1$ and $H_2$ contain magnetic anisotropy terms:
    \begin{eqnarray}
        H_1 &=& M_s - H_{2\perp} + H_{2\parallel}\cos^2\left(\varphi + \frac{\pi}{4} \right) + \frac{1}{4}H_{4\parallel}(3+\cos4\varphi)   \\
        H_2 &=& H_{4\parallel}\cos 4\varphi - H_{2\parallel}\sin 2\varphi,
    \end{eqnarray}
$H_{2\perp}$, $H_{2\parallel}$ and $H_{4\parallel}$ represent the out-of-plane uniaxial, in-plane uniaxial and in-plane biaxial anisotropy respectively, and $\varphi$ is the angle between the magnetisation vector $\mathbf{M}$ and the [100] crystallographic axis. For in-plane equilibrium orientation of $\mathbf{M}$, only the alternating in-plane angle ($\sim$m$_b$(t)/M$_s$) will lead to a rectification voltage, and we can neglect the out-of-plane component of the precession. The magnetisation precession causes a time-varying resistance change originating in the anisotropic magnetoresistance (AMR): $R(t) = R_0 - \Delta R\cos^2(\theta_{\mathbf{M-E}} + m_b(t)/M_s)$. This, together with a microwave current at the same frequency, produces a voltage, $V(t) =  I\cos(\omega t)\cdot R(t)$, and we measured the dc component which is given by $V_{\text{dc}}=(I \Delta R m_b/2M_s) \sin 2\theta_{\mathbf{M-E}}$. Using Eq. (\ref{eq:mb}) with the above approximation and focusing on the real components, we can find the dc component as:
\begin{equation}
        \text{Re}\{V_\text{dc}\} = V_\text{sym}\frac{\Delta H^2}{(H_0 - H_\text{res})^2 + \Delta H^2} + V_\text{asy}\frac{\Delta H(H_0 -  H_\text{res})}{(H_0 - H_\text{res})^2 + \Delta H^2}
        \label{eq:realVdc}
    \end{equation}
    \begin{eqnarray}
        V_\text{sym} (\theta_{\mathbf{M-E}}) &=& \frac{I \Delta R \omega }{2\gamma\Delta H (2 H_\text{res} + H_1 + H_2)} \sin(2 \theta_{\mathbf{M-E}}) h_z     \label{eq:Vsym} \\
        V_\text{asy} (\theta_{\mathbf{M-E}}) &=& \frac{ I \Delta R (H_\text{res} + H_1)}{2\Delta H (2 H_\text{res} + H_1 + H_2)} \sin(2 \theta_{\mathbf{M-E}}) (-h_x\sin\theta_{\mathbf{M-E}} + h_y\cos\theta_{\mathbf{M-E}})     \label{eq:Vasy}
    \end{eqnarray}

We used these equations to quantify $h_x$, $h_y$ and $h_z$ from the in-plane angle dependence of $V_\text{dc}$. Each FMR trace was first fit by a function with symmetric and anti-symmetric Lorentzians and both components are analysed by $V_\text{sym} (\theta_{\mathbf{M-E}})$ and $V_\text{asy} (\theta_{\mathbf{M-E}})$. In table I we list experimental measurements of the magnetisation independent in-plane and magnetisation dependent out-of-plane spin-orbit fields for our set of 8 samples.

In table II, we give the uniaxial ($H_u$) (along [1-10]) and cubic ($H_c$) anisotropies; $\mu_0M_{\text{eff}}$ and the linewidth (at a frequency of 11 GHz) for each of our 8 samples, extracted from the angle-dependent FMR measurements. In addition, we show the sample resistances and AMRs.

\newpage

\begin{center}
\begin{table}[h!]
\begin{tabular}{|c|c|c|c|c|c|c|c|c|}
  \hline
  Sample & 1 & 2 & 3 & 4 & 5 & 6 & 7 & 8 \\\hline
  Direction & [100] & [100] & [010] & [010] & [110] & [110] & [1-10] & [1-10] \\\hline
  $\mu_0h_x$ ($\mu$T) & -49 & -91 & 132 & 96 & 2 & 2 & $<$1 & $<$1 \\\hline
  $\mu_0h_y$ ($\mu$T) & -17 & -15 & -49 & -30 & 127 & 120 & -201 & -145 \\\hline
  $\mu_0h_z - \sin{\theta_{\mathbf{M-E}}}$ ($\mu$T) & 51 & 95 & -122 & -107 & 4 & 6 & 8 & 5 \\\hline
  $\mu_0h_z - \cos{\theta_{\mathbf{M-E}}}$ ($\mu$T) & 20 & 41 & 19 & 42 & 161 & 203 & -127 & -86 \\\hline
  $\mu_0h_z$ - const. ($\mu$T) & -10 & 13 & -23 & 25 & 27 & $<$1 & 27 & 2 \\\hline
\end{tabular}
\caption[S1]{Amplitudes of the spin-orbit effective fields for different directions and symmetries. }
\end{table}
\end{center}

\begin{center}
\begin{table}[h!]
\begin{tabular}{|c|c|c|c|c|c|c|c|c|}
  \hline
  Sample & 1 & 2 & 3 & 4 & 5 & 6 & 7 & 8 \\\hline
  Direction & [100] & [100] & [010] & [010] & [110] & [110] & [1-10] & [1-10] \\\hline

$\mu_0H_c$ (mT) & 59 & 66 & 61 & 65 & 62 & 62 & 60 & 58\\\hline
$\mu_0H_u$ (mT) & 59 & 43 & 45 & 68 & 40 & 38 & 51 & 65\\\hline
$\mu_0M_{\text{eff}}$ (mT) &	429 & 411 & 437 & 360 &	404 & 402 & 350 & 368\\\hline
$\mu_0\Delta H$ (mT) & 7.1 & 7.4 & 8.2 &	6.8 & 9.4 &	8.8	& 7.5 &	 6.9\\\hline
AMR ($\Omega$) & 45 &	44 & 44 &	45 &	151 &	154 & 140 & 129\\\hline
R (k$\Omega$) & 11.3 & 11.3 & 11.3 &	11.3 & 11.4 &	11.4 &	11.5 &	 10.7\\\hline
\end{tabular}
\caption {Magnetic anisotropy and transport parameters in the studied devices. }
\end{table}
\end{center}

\subsection{Theory of Intrinsic Spin-Orbit Torque}
\label{Kubo-supplementary}
The dynamical interaction of the magnetization originating from localized moments arising from the d-electrons and the
delocalized hole carriers in (Ga,Mn)As gives rise to an effective current-induced field
$\delta \mathbf{H}$.
The magnetization dynamics is then described by the Landau-Lifshitz-Gilbert
equation
\begin{equation}
  \label{eq:llg}
  \frac{d\hat{\mathbf{M}}}{dt}=-\gamma \hat{\mathbf{M}}\times (\mathbf{H}+\delta \mathbf{H})
+\alpha \frac{d\hat{\mathbf{M}}}{dt}\times \hat{\mathbf{M}}
\end{equation}
where $\mathbf{H}$ is the external and internal equilibrium effective magnetic field, $\alpha$ is the Gilbert damping
parameter and $\gamma=ge/2m_0$ is the gyromagnetic factor with $e$ the elementary charge and
$m_0$ the electron mass. The current induced field is given by%
\begin{equation}\label{eq-02}
\delta \mathbf{H}=-\frac{J_{\rm ex}}{g\mu_{\rm B}}\delta\mathbf{s},
\end{equation}
where $\mu_{\rm B}$ is the Bohr magneton, $g=2$
corresponds to the localized  $d$-electrons in (Ga,Mn)As, and
 $J_{\rm ex}=55$ meV nm$^{3}$ is the antiferromagnetic kinetic-exchange coupling between the localized d-electrons and the
valence band holes, termed $J_{pd}$. $\delta\mathbf{s}$ is the current induced non-equilibrium
spin densities. We model the carriers in these  systems are modeled  by a  Hamiltonian with a kinetic exchange coupling term
$
  H=H_{\rm GaAs}+H_{\rm ex}$, where $H_{\rm ex}=J_{\rm ex}c_{\rm Mn} S_{\rm Mn} \hat{\mathbf{M}}\cdot{\mathbf s}$,
$H_{\rm GaAs}$ refers to the 4-band strained Kohn-Luttinger Hamiltonian for the hole systems of GaAs (see below),
$\mathbf s$ is the $4\times 4$ spin operator for
the holes described by the four-band Kohn-Luttinger model, $S_{\rm Mn}=5/2$, and $c_{\rm Mn}$ corresponds to the
Mn local spin-density.

The current-induced spin density has two contributions,
$ \delta\mathbf{s}=\delta\mathbf{s}^{\rm ext}+\delta\mathbf{s}^{\rm int} $.
  The extrinsic contribution, $\delta\mathbf{s}^{\rm ext}$,\cite{Manchon:2009_a,Chernyshov:2009_a}
  arises from the non-equilibrium steady state distribution function of the carriers
  due to the interaction of the applied electric field and the spin-orbit coupling(SOC) carriers, {\it i.e.} predominantly independent of the
  magnetization and therefore of field-like form. However, there is another contribution not
  discussed theoretically before which is the focus of our study. This contribution
   arises from the electric-field induced polarization of the spins as they accelerate between
  scattering events, {\it i.e.} of purely intrinsic origin arising from the band structure of the system, which has
  the form, $\delta\mathbf{s}^{\rm int} \propto \hat{M}\times \mathbf{a}(\mathbf{E})$, where $\mathbf{a}(\mathbf{E})$
  is an in-plane function linear in the electric field that depends on the symmetry of the SOC
  responsible for the effect, Rashba or Dresselhaus, as discussed in the main text. This gives rise to an anti-damping
  torque, ${\mathbf{ \tau}}_{\rm anti-damp}\propto \hat{M}\times(\hat{M}\times \mathbf{a}(\mathbf{E}))$ and it is, in the case of
  (Ga,Mn)As, of the same order of magnitude as the extrinsic field-like SOT.

 This current-induced non-equilibrium
spin densities, $\delta\mathbf{s}$, can be calculated by
the linear Kubo response theory:\cite{Garate:2009_a}
\begin{equation}\label{eq:kubo_cif}
   \delta\mathbf{s} = \frac{\hbar }{2\pi V} {\rm Re} \sum_{\mathbf{k},a,b}
   (\mathbf{\sigma})_{ab} (e\mathbf{E}\cdot \mathbf{v})_{ba}
   [G^A_{\mathbf{k}a}G^R_{\mathbf{k}b}-G^R_{\mathbf{k}a}G^R_{\mathbf{k}b}],
\end{equation}
where the Green's functions
$G^R_{\mathbf{k}a}(E)|_{E=E_F}\equiv G^R_{\mathbf{k}a}=1/(E_F-E_{\mathbf{k}a}+i\Gamma)$,
with the property $G^A=(G^R)^*$.
The carrier states are labeled by momentum $\mathbf{k}$,
band index $a$, and $E_F$ is the Fermi energy.
$\Gamma=\hbar/2\tau$ is the spectral broadening
corresponding to a relaxation time $\tau$.
Here the matrix elements  of an operator $\hat{C}$ are
$(\hat{C})_{ab}\equiv \left< \mathbf{k}a \right| \hat{C} \left| \mathbf{k}b\right>$
or $(\hat{C})_{a}\equiv \left< \mathbf{k}a \right| \hat{C} \left| \mathbf{k}a\right>$.
The intra-band contributions in the above expressions correspond to the  component already discussed before
which gives rise to the field-like torque,\cite{Manchon:2009_a,Chernyshov:2009_a,Garate:2009_a} and the inter-band contribution is the one that gives rise to the intrinsic anti-damping SOT in analogy
to the intrinsic SHE.

The expression for $\delta\mathbf{s}^{\rm int}$ in the clean limit is given by
\begin{eqnarray}
\delta\mathbf{s}^{\rm int} & = &
  \frac{\hbar }{V}
  \sum_{\mathbf{k},a\neq b}
  \frac{ \Ima[(\mathbf{s})_{ab}(e\mathbf{E}\cdot\mathbf{v})_{ba}]}{(E_{\mathbf{k}a}-E_{\mathbf{k}b})^2}
  (f_{\mathbf{k}a}-f_{\mathbf{k}b}).
\end{eqnarray}
Here $f_{\mathbf{k}a}$ are the Fermi-Dirac distribution functions corresponding to band
energies $E_{\mathbf{k}a}$.
In the presence of disorder, as it is the case for (Ga,Mn)As, the resulting expression
are approximated by
\begin{eqnarray}\nonumber
  \delta \mathbf{s}^{\rm int} & = &
  \delta \mathbf{s}^{(1)}+\delta\mathbf{s}^{(2)}
  \\ \nonumber
  \delta \mathbf{s}^{(1)} & = &
  -\frac{1}{ V}
  \sum_{\mathbf{k},a\neq b}
  2 \Rea[({\bm \sigma})_{ab}(e\mathbf{E}\cdot\mathbf{v})_{ba}]
  \\
  &\times & \frac{\Gamma (E_{\mathbf{k}a}-E_{\mathbf{k}b})}{[(E_{\mathbf{k}a}-E_{\mathbf{k}b})^2+\Gamma^2]^2}
  (f_{\mathbf{k}a}-f_{\mathbf{k}b})
\label{eq:semic_inter}
\\ \nonumber
  \delta\mathbf{s}^{(2)} & = &
 - \frac{1}{ V}
  \sum_{\mathbf{k},a\neq b}
  2 \Ima[({\bm \sigma})_{ab}(e\mathbf{E}\cdot\mathbf{v})_{ba}]
  \\ \nonumber
  &\times &
  \frac{\Gamma^2- (E_{\mathbf{k}a}-E_{\mathbf{k}b})^2}{[(E_{\mathbf{k}a}-E_{\mathbf{k}b})^2+\Gamma^2]^2}
  f_{\mathbf{k}a}.
\end{eqnarray}
Here we have ignored small numerical corrections due to the $G^R_{\mathbf{k}a}G^R_{\mathbf{k}b}$ terms which
can be shown to formally  vanish in a weak disorder situation and whose rapid oscillations can lead to numerical instabilities
giving rise to systematic errors.

The hole-valence system is described by
 $ H_{\rm GaAs}=H_{\rm KL}+H_{\rm strain}$,
where the first term is the Kohn-Luttinger Hamiltonian and the second contains the strain effects.
The four-band Kohn-Luttinger Hamiltonian in the hole-picture is
\begin{eqnarray}
  \label{eq:hkl}
  H_{\rm KL} & = &
  \dfrac{\hbar^2k^2}{2m_0}
  \left( \gamma_1+\frac{5}{2}\gamma2 \right){\bf I}_4-
  \dfrac{\hbar^2}{m_0}\gamma_3 \left(\mathbf{k}\cdot\mathbf{J} \right)^2 \\
& + &     \dfrac{\hbar^2}{m_0}(\gamma_3-\gamma_2)
    \left(k_x^2J_x^2+k_y^2J_y^2+k_z^2J_z^2 \right).\nonumber
\end{eqnarray}
Here, $\mathbf{k}$ is the momentum of the holes, $m_0$ is the electron
mass, $\gamma_1 = 6.98$, $\gamma_2 = 2.06$,
and $\gamma_3 = 2.93$ are the Luttinger parameters, ${\bf I}_4$ is the
$4\times 4$ identity matrix and $\mathbf{J}=(J_x,J_y,J_z)$ are the
$4\times 4$ angular momentum matrices of the holes.
Here the hole spin $\mathbf{s}={\bm J}/3$,
where ${\bm s}$ are the spin matrices for holes.\cite{Abolfath:2001_a}

The strain Hamiltonian in the hole-picture is
\begin{eqnarray}
  \label{eq:c4strain}
  H_{\rm strain} & = & b
  \left[
    \left( J_x^2-\frac{\mathbf{J}^2}{3}\right) \epsilon_{xx}+
    {\rm c.p.}
  \right]  \nonumber \\
&& -    C_4\left[
    J_x\left(\epsilon_{yy}-\epsilon_{zz}\right)k_x+{\rm c.p.}
  \right] \\
&& -  C_5
  \left[
    \epsilon_{xy}(k_yJ_x-k_xJ_y)+{\rm c.p.}
  \right] ,\nonumber
\end{eqnarray}
where $\epsilon_{ij}$ is the strain tensor and $b=-1.7$ eV is the axial deformation
potential. $C_4$ is the magnitude of the momentum-dependent
Dresselhaus-symmetric strain term and $C_5$ is the magnitude of the
Rashba-symmetric strain term. In our calculations, we use the value
$C_4 = 10$ eV${\rm \AA}$ calculated\cite{Silver:1992_a,Stefanowicz:2010_a} from first principles for holes in
(Ga,Mn)As and $C_5 = C_4$. To the best of our knowledge, there is no
measurement or calculation for the $C_5$ term in (Ga,Mn)As.
In our calculations we set $\gamma_2=\gamma_3$ within the spherical approximation and for the parabolic approximation
we set $\gamma_2=\gamma_3=0$ and take $\gamma_1=2$.
The external electric field magnitude is set to
$E=0.02$ mV/nm (from the experimental values), the disorder broadening
to $\Gamma=25$ meV, and the strain to $\epsilon_{xx}=\epsilon_{yy}=-1.1 \epsilon_{zz}=-0.3\%$ and $\epsilon_{xy}=-0.15\%$.
The first term of the strain Hamiltonian is momentum independent.
{\it The other two terms are momentum-dependent and they are
 essential for the generation of  SOT because they break the space inversion symmetry.}
The second term has a Dresselhaus symmetry and the third has a Rashba symmetry.
As described in the experimental results, these symmetries are shared by the observed SOT.
In this discussion we have neglected cubic Dresselhaus terms, allowed by the GaAs
symmetry, since the experimentally observed SOTs vary linearly with strain.\cite{Chernyshov:2009_a,Fang:2010_a}


\end{appendix}

\end{document}